# Van der Waals Stacking Induced Topological Phase Transition in Layered Ternary Transition Metal Chalcogenides

Junwei Liu,[1,†] Hua Wang,[2,†] Chen Fang,[3] Liang Fu,[1] Xiaofeng Qian[2]*

[1] Department of Physics, Massachusetts Institute of Technology, Cambridge, Massachusetts 02139, USA

[2] Department of Materials Science and Engineering, Texas A&M University, Dwight College of Engineering, College Station, Texas, 77843, USA

[3] Institute of Physics, Chinese Academy of Sciences, Beijing 100190, China

† These authors contributed equally to this work.

*Correspondence and requests for materials should be addressed to X.Q.
(email: feng@tamu.edu).

**Abstract**

Novel materials with nontrivial electronic and photonic band topology are crucial for realizing novel devices with low power consumption and heat dissipation, and quantum computing free of decoherence. Here using first-principles approach, we predict a class of ternary transition metal chalcogenides (TTMC) MM'Te$_4$ exhibits dual topological characteristics: quantum spin Hall (QSH) insulators in their 2D monolayers and topological Weyl semimetals in their 3D noncentrosymmetric crystals upon van der Waals (vdW) stacking. Remarkably, we find that one can create and annihilate Weyl fermions, and realize the transition between Type-I and Type-II Weyl fermions by tuning vdW interlayer spacing. Our calculations show that they possess excellent thermodynamic stability and weak interlayer binding, implying their great potentials for experimental synthesis, direct exfoliation and vdW heterostacking. Moreover, their ternary nature will offer more tunability for electronic structure by controlling different stoichiometry and valence charges. Our findings provide an ideal materials platform for realizing QSH effect and exploring topological phase transition, and will open up a variety of new opportunities for two-dimensional materials and topological materials research.

**Introduction**

The seminal discovery of quantum spin Hall (QSH) effect (*1-6*) engendered a new chapter of topological materials research in condensed matter physics and materials science (*7, 8*), followed by the discoveries of three-dimensional (3D) topological insulator (*9-13*), quantum anomalous Hall insulator (*14, 15*), topological crystalline materials (*16-20*), Weyl semimetals (*21-28*), etc. An interesting generic aspect is the presence of special surface/edge states that are topologically protected upon small perturbations, hence inelastic scattering induced heat dissipation is minimized. In contrast, conventional electronics suffers from severe local heating as any structural or chemical defect could introduce additional scattering and reduce carrier transmission. These topological phases, if materialized and integrated at the device level, could be advantageous for many novel low-power and low-dissipation electronic applications. Along with materials discoveries with individual topological phases, there is an increasing interest to understand topological phase transition, for examples, from nontrivial to trivial topology and from one nontrivial phase to another (*7, 8*) through chemical doping, elastic strain engineering, electric and magnetic field perturbations, etc.

In parallel, 2D transition metal dichalcogenides (TMDCs) also arouse substantial interest (*29-31*) in their distinct properties, for example, extraordinarily enhanced excitonic photoluminescence in monolayer 1H-MoS$_2$.(*32*) Most TMDCs have trivial topology, either metallic or semiconducting with large gap. However, among their polymorphic variants, 1T structure of binary TMDCs such as MoTe$_2$ and WTe$_2$ is subject to the Peierls instability, resulting in a dimerized 1T' structure in their monolayers, which surprisingly leads to a variety of novel quantum phenomena such as nontrivial electronic topology (*33, 34*) and nonsaturating giant magnetoresistance (*35*). In particular, their monolayers were predicted to exhibit QSH effect (*33*), and the weak vdW coupling between different monolayers makes them suitable to realize multi-functional materials/devices by stacking with other 2D materials (*36*), such as topological field transistors (*33*). Encouragingly, there are great experimental progresses of nontrivial 2D TMDCs. For example, Peierls-distorted MoTe$_2$, though metastable, has been successfully synthesized in experiments, and recently characterized with a fundamental gap of about 60 meV by both optical and electrical measurements (*37-39*). Nonetheless, special encapsulation and assembly are required to prevent them from transforming into the more stable but topologically trivial 1H phase. Meanwhile, there are limited choices within binary TMDCs for the discovery and design of materials with nontrivial topology. Therefore, for the ease of experimental realization and device implementation, it is highly desirable to go beyond the present binary $MX_2$ materials and explore other possibilities for large gap topological materials that are thermodynamically stable.

Here we report our theoretical predictions of four layered ternary transition metal chalcogenides (TTMC), namely MM'Te$_4$ with M=(Nb, Ta) and M'=(Ir, Rh), as a new class of topological materials which displays QSH effect with sizable fundamental gaps in their monolayers, become topological Weyl semimetals in their bulk form. Remarkably, the vdW interlayer interaction drives topological phase transition via formation and splitting of Dirac points and the latter results in the emergence of topological Weyl nodes and the smooth transition between Type-I and Type-II Weyl fermions. All four ternary compounds are found to be thermodynamically and dynamically stable, and their interlayer binding energies are slightly weaker than that of graphite, making them promising for experimental synthesis, mechanical exfoliation, vdW heterostructuring, and direct electronic characterization thereafter, thereby providing an ideal materials platform for exploring both novel device applications and fundamentals of topological phase transitions.

**Results**

Bulk ternary transition metal chalcogenides MM'Te$_4$ are noncentrosymmetric orthorhombic layered materials (*40, 41*) with space group Pmn2$_1$ (commonly known as T$_d$ polymorph) as shown in Fig. 1, where M=(Nb, Ta) of Group 5, and M'=(Ir, Rh) of Group 9. Different from their bulk, monolayers MM'Te$_4$ are centrosymmetric 2D semiconductors with space group P2$_1$/m (known as 1T' polymorph) whose inversion center is located at the middle of two adjacent M atoms or M' atoms along the dimerization direction (+y). The atomic structures were fully relaxed using first-principles density functional theory (DFT) with interlayer vdW interactions taken into account. Taking NbRhTe$_4$ as an example, each layer of bulk T$_d$-NbRhTe$_4$ is similar to monolayer 1T' TMDCs, *i.e.* exhibiting Peierls distortion induced dimerization along y with zig-zag quasi-one dimensional chains formed along x. Compared to the 1T' TMDCs, the unit cell along the dimerization direction doubles as the two metallic elements repeat as [--M'-M--M-M'--]. Here, single and double dashes denote short and long bond between adjacent metallic atoms. This doubling is due to different electronegativities of M and M', resulting in charge transfer between M and M', hence stronger bonding and shorter bond length in M-M' than M--M or M'--M' bond. Consequently, it is energetically more stable than the structure with direct [–M'–M–] repeating units.

Without SOC two massless Dirac cones (Λ) show up along the RYR high symmetry line on the Brillouin zone edge indicated in Fig. 2A and 2B, which implies the presence of band inversion around Y. Once SOC is turned on, a band gap of 65 meV is opened up at the Dirac points while time reversal symmetry (TRS) is still preserved. The corresponding band structures for all four monolayer MM'Te$_4$ are shown in fig. S1. Band inversion at Y together with TRS renders the characteristic of quantum spin Hall insulator. To verify its topology, we calculate the $Z_2$ topological index using two independent methods including the parity product at time-reversal invariant momenta(*42*) and the Berry-phase based n-field method(*43*), and both methods give $Z_2$ index of 1, thereby confirming the nontrivial topology. The nontrivial topology is also originated from the Peierls distortion in 1T' structure which is indeed very similar to 1T' binary TMDCs, while the microscopic mechanism of band inversion is a little different. Here both the top valence band and bottom conduction band near the Fermi level of monolayer NbRhTe$_4$ are dominated by $dx^2$-$y^2$, dyz, and $dz^2$ of Nb atoms (see fig. S8), corresponding to a d-d band inversion that is different from the p-d band inversion in 1T' binary TMDCs. Furthermore, as all Nb atoms are located in the middle of the three atomic layers, their $Z_2$ topology should be more robust against local environment, such as vdW heterostacking(*36, 44*). More detailed analysis can be found in the Supplementary Materials.

Given a piece of topologically nontrivial and large enough 2D monolayer MM'Te$_4$ whose edge is interfaced with a trivial material such as vacuum, there will be an unavoidable nontrivial-to-trivial topological phase change across the interface. It guarantees the appearance of two counter-propagating helical states along the edges of 2D host materials in the absence of magnetic field or non-TRS breaking impurities, connecting the bulk valence states to the conduction states. To investigate the edge states, we constructed first-principles tight-binding (TB) Hamiltonians for each monolayer MM'Te$_4$ and applied Green's function method to calculate the electronic structure of semi-infinite monolayer MM'Te$_4$ from which both bulk density of states (DOS) and edge DOS are extracted. Figure 2C and 2D present the corresponding bulk and edge DOS as well as spin ⟨$S_z$⟩-resolved edge DOS for the y and x edges of monolayer NbRhTe$_4$, respectively. Both unambiguously reveal the fully gapped bulk electronic band structures and the presence of nontrivial edge states. Results for the other three monolayer MM'Te$_4$ are shown in fig. S2. The spin-resolved edge DOS clearly demonstrates the opposite spin carried by two counter-propagating states which cross each other at time-invariant momenta of 1D Brillouin zone, that is, $\bar{\Gamma}$ and $\bar{Y}$ along the y edge and $\bar{\Gamma}$ and $\bar{X}$ along the x edge guaranteed by TRS. In contrast to conventional transport channels that are subject to severe scattering upon structural impurities and defects, these edge states are topologically protected and thus free of scattering due to nonmagnetic defects and impurities, making them ideal for low dissipation electronic applications.

Bulk $T_d$-MM'Te$_4$ are formed by vdW bonded MM'Te$_4$ layers. In principle, if the interlayer interaction is weak, directly stacked multilayer MM'Te$_4$ could exhibit even-odd oscillation in their $Z_2$ topological index, and essentially form 3D weak topological insulators. However, the interaction between layers is not as weak as one would think, and the stacking geometry is not as simple as the AA stacking which is obvious according to the symmetry operation of their Pmn2$_1$ space group.

The absence of inversion symmetry together with the non-negligible interlayer interaction introduces a topological phase transition of bulk $T_d$-MM'Te$_4$ into Weyl semimetal phase. Weyl points are essentially "magnetic monopoles" with topological charges defined by Chern number, i.e., integration of Berry curvature enclosing a single Weyl point divided by $2\pi$. It can be either +1 or -1, corresponding to the outgoing and incoming effective magnetic flux, respectively. We performed a detailed scan of bulk electronic band structure to locate the intersecting points between the top valence band and the bottom conduction band, and found all four materials possess at least four Weyl points. Here, as illustrated in Fig. 3A, bulk $T_d$-NbRhTe$_4$ has eight Weyl points, which can be divided into two groups with their energy-momentum positions $(E, k_x, k_y, k_z)$ of (0.125eV, $\pm 0.16$, $\pm 0.12$, 0) and (-0.066eV, $\pm 0.01$, $\pm 0.07$, 0), respectively, where momenta are given in the fractional reciprocal lattice coordinate and the two points in the first quadrant are denoted by $W_1$ and $W_2$. All the Weyl points are locked on the $k_z$=0 plane, which is due to their intrinsic Pmn2$_1$ space group as discussed later. Moreover, all of them are well separated in the momentum space, making it much easier for experimental detection. Correspondingly, the band structures near $W_1$ and $W_2$ are plotted in Fig. 3B and 3C. Constant energy planes of 0.125 eV and -0.066 eV cut through the top valence and bottom conduction bands of $W_1$ and $W_2$, respectively, forming electron and hole pockets, which indicates they are Type-II Weyl points(*34*). The calculated Chern numbers enclosing $W_1$ and $W_2$ are -1 and +1, implying negative and positive "magnetic monopole", i.e., effective net Berry flux going inward and outward, respectively. The corresponding Berry curvatures are shown in Fig. 3D and 3E which are highly anisotropic due to the vdW stacking nature of bulk $T_d$-MM'Te$_4$.

A unique feature of Weyl semimetals is the presence of exotic Fermi arcs on surfaces, terminated at the projections of a pair of Weyl points with opposite Chern numbers, which is indeed confirmed in our calculations as shown in Fig. 4. All energy contours of the bulk states (Fig. 4A, 4B, 4E and 4F) are normal closed loops, while for the surface states (Fig. 4C, 4D, 4G and 4H) there are some nontrivial open Fermi arcs ending exactly at the surface projections of the bulk Weyl points with opposite Chern numbers marked by red and blue circles. Moreover, some Fermi arcs spread a very large region of surface BZ since the Weyl points are very well separated, which makes experimental detection such as angle-resolved photoemission emission spectroscopy much easier. Bulk and surface DOS for all four bulk materials are shown in the fig. S3-6.

The above results demonstrate that monolayer 1T'-MMTe$_4$ is a 2D TI, and bulk $T_d$-MM'Te$_4$ is a Weyl semimetal. Since the former is the decoupled layers limit of the latter with AA stacking, it is important to understand the evolution of topological phase transition between these two cases as the layer separation decreases. Direct first-principles calculations will be too heavy. Nonetheless, this can be achieved by using a linear interpolation between two effective tight-binding Hamiltonians, one with a large lattice constant c (18.46Å used here for $T_d$-NbRhTe$_4$) corresponding to the decoupled layers limit and one with the bulk Td structure (c=13.46Å). The linear interpolation simulates the evolution of the layer separation. Subsequently, one can determine the whole evolution process of Weyl points by scanning the $k_z$

=0 plane of Brillouin zone, locating the Weyl points if any, and verifying their topology by computing the associated Chern numbers.

Physically, the interlayer coupling plays two roles: (i) it breaks the inversion symmetry and splits the spin degeneracy of all bands and (ii) it modifies the band gap near X significantly and causes band inversions nearby. Each inversion of two non-degenerate bands in a time-reversal invariant system leads to the creation or annihilation of four Weyl points(*21*).

Figure 3F shows the results of topological phase transitions for $T_d$-MM'Te$_4$ with lattice constant c approaching the equilibrium value, with color changing from purple to blue as c decreases. Four groups of Weyl points are clearly seen in Fig. 3F. The first group of Weyl points initially appear at relatively large lattice constant (purple-to-blue dots) near Y when two Dirac points on YΓY are split into four Weyl points which quickly evolve towards YR, and then annihilate with decreasing c. The second group also appears on ΓY but close to Γ, and eventually annihilate on ΓX as well. The third group emerges on YR close to the original two massive Dirac points of monolayer 1T'-MMTe$_4$, then moves towards ΓY, but they do not reach any high-symmetry line at the equilibrium c, leaving four Weyl points, where the one in the first quadrant has +1 Chern number. The last group emerges on ΓX, then moves away from ΓX, leaving four Weyl points at the equilibrium lattice constant c, where the one in the first quadrant has -1 Chern number. As a result, there are total eight Weyl points left in the BZ, exactly the same as the ones illustrated in Fig. 3A for the relaxed bulk $T_d$-MM'Te$_4$. It is worth noting that during the above topological evolution, both Type I and Type II Weyl points show up which are marked in the first quadrant of Fig. 3F. Moreover, the third group starts with Type I Weyl points, but ends up with Type II Weyl points at equilibrium c, indicating a dramatic band structure evolution as well.

This also implies one can induce the transition between Type I and Type II Weyl points by external perturbations such as elastic strain.

This process involves a series of topological phases transitions sharing some common features summarized here (see Supplementary Materials for more details). First, all the Weyl points are pinned to the $k_z = 0$ plane due to the symmetry $T * M_x * G_y$. Second, two Weyl points emerge from some high-symmetry line and move towards another, perpendicular high-symmetry line. Third, the layer coupling in equilibrium $T_d$-MM'Te$_4$ is not very small, as some Weyl points traverse almost half of the BZ to meet and annihilate. Finally, each creation or annihilation of a pair of Weyl points along a high-symmetry line is accompanied by the change of the topological $Z_2$-index on the plane in the BZ spanned by that line and $k_z$-axis. Due to the third point above, one does not expect to recover the whole process using an effective $k \cdot p$ model based at Y. However, in the Supplementary Materials, we show that the creation and the annihilation of the first group of Weyl points, *i.e.*, when c is still large, can be captured by a simple model.

For the experimental synthesis and mechanical exfoliation, it is crucial to check the related materials properties such as dynamic and thermodynamic properties as well as interlayer binding energy. Figure 5A shows the phonon dispersion of monolayer NbRhTe$_4$ calculated by first-principles density-functional perturbation theory, while figure S7 presents the results for all four materials. Negative frequency is absent in all four monolayer cases, suggesting excellent dynamic stability upon small structural perturbation. Furthermore, we calculated the total energy difference between the fully relaxed 1H and 1T' structures. The results shown in Fig. 5B reveals that 1T' structure is more stable than 1H structure by 0.53, 0.36, 0.43 and 0.26 eV/formula unit (f.u.) for TaIrTe$_4$, TaRhTe$_4$, NbIrTe$_4$, and NbRhTe$_4$, respectively. It is thus very promising for direct experimental synthesis. In addition, the total energy difference between bulk $T_d$-MM'Te$_4$ and monolayer 1T'-MM'Te$_4$ displayed in Fig. 5C yields the interlayer binding energy ($E_b$) of 25, 26, 26, and 27 meV/Å$^2$, respectively. All of them are even less than that of graphite (~30meV/Å$^2$ using the same PBE exchange-correlation functional and optB88-vdW correlation functional), validating the feasibility of direct mechanical exfoliation to obtain their multi-layer or monolayer structures. Lastly, as shown in Fig. 5D, all four 2D monolayer MM'Te$_4$ have sizable fundamental gaps of 23, 58, 30, and 65 meV for TaIrTe$_4$, TaRhTe$_4$, NbIrTe$_4$, and NbRhTe$_4$, respectively. In summary, the above results demonstrate that Peierls-distorted MM'Te$_4$ are both thermodynamically and dynamically stable and their interlayer binding energy is relatively weak, therefore particularly promising for experimental synthesis and mechanical exfoliation whose finite fundamental gap shall allow direct experimental measurements.

## Discussion

Among the four predicted topological materials, T$_d$-polytypical NbIrTe$_4$ and TaIrTe$_4$ were synthesized and characterized with the Pmn2$_1$ space group two decades ago (*40, 41*) and the measured electrical resistivity of bulk T$_d$-NbIrTe$_4$ and T$_d$-TaIrTe$_4$ increases with increasing temperature. These experiments provide a solid support for our theoretical results on thermodynamic and dynamic stabilities and semi-metallic electronic structure. It is therefore highly plausible that two other materials NbRhTe$_4$ and TaRhTe$_4$ may be synthesized as well. Moreover, it is worth to note that two metal elements M and M' of MM'Te$_4$-based TTMCs possess different valence charges, making them distinctly different from convectional binary TMDCs and their alloys such as WTe$_2$ and W$_{1-x}$Mo$_x$Te$_2$. This ternary nature shall allow one to adjust the stoichiometric ratio of M and M', thereby providing another viable strategy to manipulate the Fermi level. It will permit one to fine-tune the Fermi level position with respect to Weyl points, enabling a facile realization of chiral anomaly. Furthermore, different choices of metallic elements in TTMCs can significantly enlarge the discovery and design space for topological materials. Finally, similar atomic and electronic structures as well as the nontrivial 2D and 3D topologies between these ternary compounds and $WTe_2$ strongly suggest that they may also exhibit giant magnetoresistance, which is highly worth of further experimental exploration.

It is important to emphasize that this new class of TTMC MM'Te$_4$ predicted here can simultaneously serve as two-dimensional QSHIs in their 2D monolayers and Weyl semimetals in their 3D bulk crystals. The Peierls-distorted monolayer 1T' or bulk $T_d$ ternary structure may be realized in a more general formula with M=(Nb, Ta) and M'=(Ru, Os, Rh, Ir), providing additional materials and alloying/doping choices, meanwhile, the weak interlayer binding attraction allows them for direct mechanical exfoliation and vdW heterostructuring (*36, 44*). Such rich topologies and wide possibilities as well as experimental flexibility may empower them as a unique configurable and tunable platform for studying topological phase transition in these ternary vdW layers and developing novel quantum electronics with low power consumption/heat dissipation and quantum computing free of decoherence.

In conclusion, the present findings extend the topological materials study to a new paradigm with a much larger search and design space. It will open up a variety of unprecedented research opportunities for not only two-dimensional layered materials research (*29-31*), but also the rapidly growing quest for stable topological materials that are easy for experimental synthesis and characterization as well as topological device implementation.

**Materials and Methods**

Ground-state atomic structures of monolayer 1T' and bulk $T_d$-MM'Te$_4$ were calculated by first-principles density functional theory (DFT) (*45, 46*) calculations implemented in the Vienna

Ab initio Simulation Package (VASP) with plane wave basis (*47*) and the projector-augmented wave method (*48*). We adopted the Perdew-Berke-Ernzerhof (PBE)'s form (*49*) of exchange-correlation functional within the generalized-gradient approximation (GGA) (*50, 51*) and the Monkhorst-Pack (*52*) **k**-point sampling for the zone integration. More specifically, an energy cutoff of 400 eV and Monkhorst-Pack **k**-point sampling of 12×3×3 for bulk and 12×3×1 for monolayer were chosen to fully relax the atomic structures with maximum residual force less than ~ 0.01 eV/Å. The interlayer van der Waals (vdW) interactions (*53*) were taken into account by applying the optB88-vdW correlation functional proposed by Klimeš et al. (*54*) To confirm the dynamical stability of the relaxed structures, phonon band structures were calculated by using density-functional perturbation theory (*55*).

To investigate the topological nature of MM'Te$_4$, we first transformed the Kohn-Sham eigenstates of DFT calculations into a set of highly localized quasiatomic orbitals accompanied with first-principles tight-binding Hamiltonian (*56, 57*). The latter was used for the calculation of topological indices such as $Z_2$ topological index and Chern number. More specifically, we computed the $Z_2$ topological index of monolayer materials using two independent methods, including the parity product at time-reversal invariant momenta (*42*) and the Berry-phase based n-field method (*43*). For bulk $T_d$-MM'Te$_4$, we first performed a coarse scan to locate the rough regions of potential Weyl points between the top valence band and the bottom conduction band, and then used simplex search method (*58*) to find their exact locations with an energy tolerance of $10^{-10}$ eV. Subsequently, we constructed a triangulated sphere near each Weyl point and calculated the corresponding Berry curvature and Chern number. Here the radius of sphere was chosen to be small enough to avoid two Weyl points within a single sphere. Bulk, edge and surface DOS are calculated using the imaginary part of the retarded Green's function (*59*) of semi-infinite atomistic models of 2D and 3D MM'Te$_4$. A kinetic energy cutoff of 300 eV and a Monkhorst-Pack **k**-point sampling of 8×3×3 for bulk MM'Te$_4$ and 8×3×1 for monolayer MM'Te$_4$ were used for the above electronic structure calculations.

**Acknowledgments**

**Funding:** X.Q. and H.W. acknowledge the start-up funds from Texas A&M University and Texas A&M Supercomputing Facility for providing supercomputing resources. J.L. was supported by the MRSEC Program of the National Science Foundation under award number DMR-1419807. C.F. was supported by the National Thousand-Young-Talents Program of China. L.F. was supported by the DOE Office of Basic Energy Sciences, Division of Materials Sciences and Engineering under Award No. DE-SC0010526.

**Author contributions:** X.Q. conceived and supervised the project. X.Q., J.L., H.W. and C.F. carried out the calculations. X.Q., J.L., H.W., C.F. and L.F. analyzed the results. X.Q. wrote the initial draft, and J.L., C.F., L.F. and H.W. contributed to manuscript revision.

**Competing interests:** The authors declare that they have no competing interests.

**Figures and Tables**

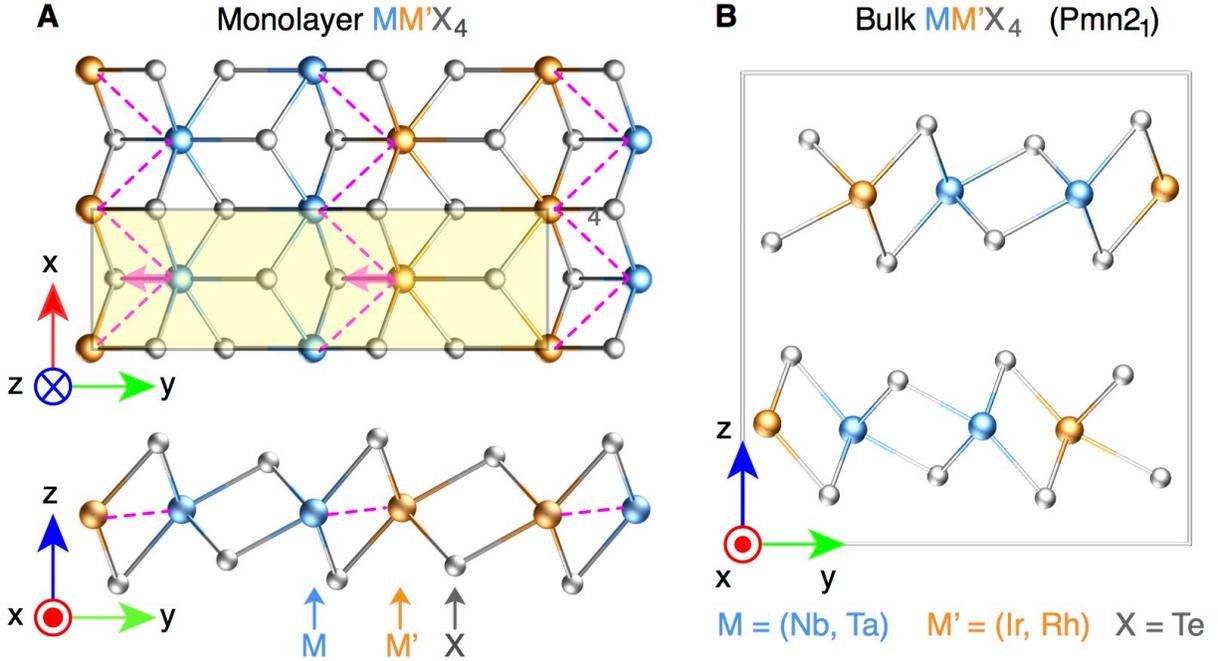

**Fig. 1. Atomic configuration of ternary transition metal chalcogenides MM'Te$_4$.** (**A**) Top and side view of two-dimensional MM'Te$_4$ in the monolayer 1T' structure. Dashed purple lines indicate in-plane zig-zag metallic bond along the x direction owing to Peierls distortion in the y direction. (**B**) Side view of three-dimensional MM'Te$_4$ in their bulk T$_d$ structure. Both M and M' are transition metal atoms. M, M', and Te are indicated by blue, orange, and gray atoms. M stands for Ta or Nb, M' stands for Ir or Rh in the present work.

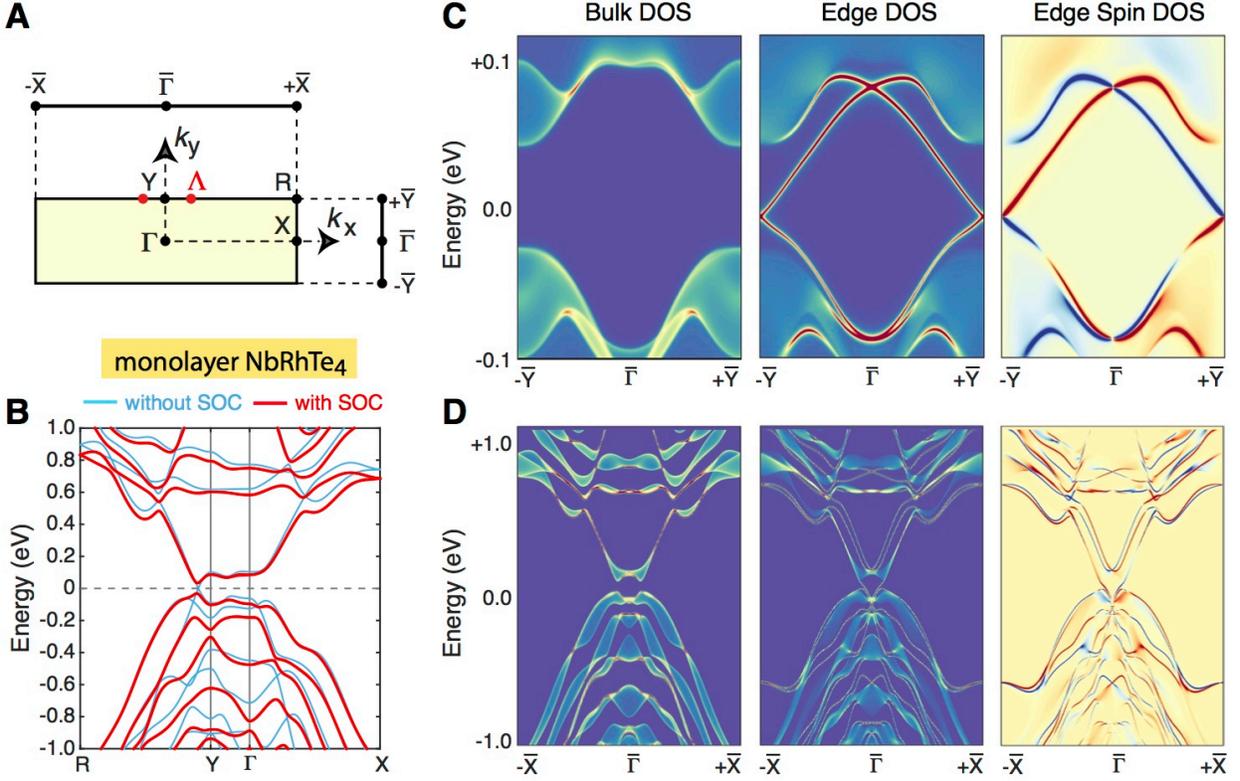

**Fig. 2. Electronic structure of quantum spin Hall insulators 1T'-MM'Te$_4$ using monolayer 1T'-NbRhTe$_4$ as a representative case.** (**A**) 2D Brillouin zone of monolayer 1T' NbRhTe$_4$ and its projections onto the $k_x$ and $k_y$ edges. Dirac points "Λ" in the absence of spin-orbit coupling are indicated by red dots along the YRY zone boundary. (**B**) Electronic band structure of monolayer NbRhTe$_4$ with and without spin-orbit coupling indicated by red and blue lines. (**C, D**) Bulk DOS, edge DOS, and spin <$S_z$>-resolved edge DOS of monolayer NbRhTe$_4$ along the $k_y$ and $k_x$ edge, respectively. Red and blue color in the edge spin DOS indicate the positive and negative spin, respectively.

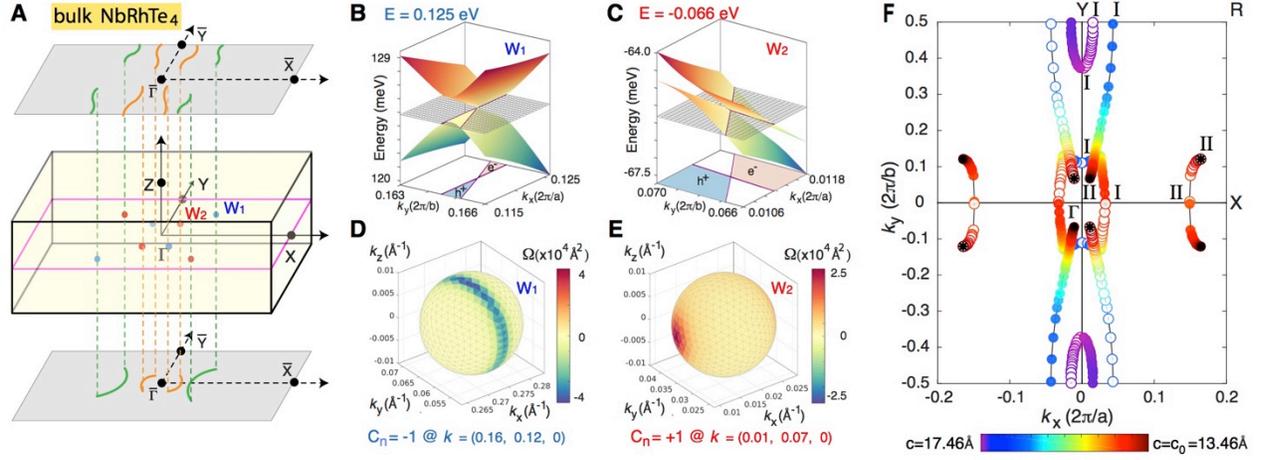

**Fig. 3. Electronic structure and topological phase transition of Weyl semimetal bulk $T_d$-NbRhTe$_4$.** (**A**) 3D Brillouin zone and its projections onto top and bottom surfaces along the z direction. Eight gapless Weyl points are located in the $k_z=0$ plane, indicated by red and blue dots (i.e. Chern number of +1 and -1). "W$_1$" and "W$_2$" represent the two Weyl points in the first quadrant of $k_z=0$ plane. Their projections on top and bottom (001) surfaces form Fermi arcs passing through individual pairs of Weyl points. (**B,C**) Band structure close to W$_1$ and W$_2$, respectively. Conduction and valence bands intersect with a constant energy plane (indicated by a gray mesh), forming a linear cone and separating electron and hole pockets. (**D,E**) Berry curvature $\Omega$ around Weyl points W$_1$ and W$_2$ with high anisotropy, respectively. (**F**) Topological evolution of Weyl points locked on the $k_z=0$ plane with decreasing effective interlayer distance. Solid dots: Chern number of +1; empty dots: Chern number of -1. Type I and Type II Weyl points are labeled at the beginning and the end of the evolution trajectories in the first quadrant. Final eight Weyl points at equilibrium $c$ are marked by dark asterisk symbols "*".

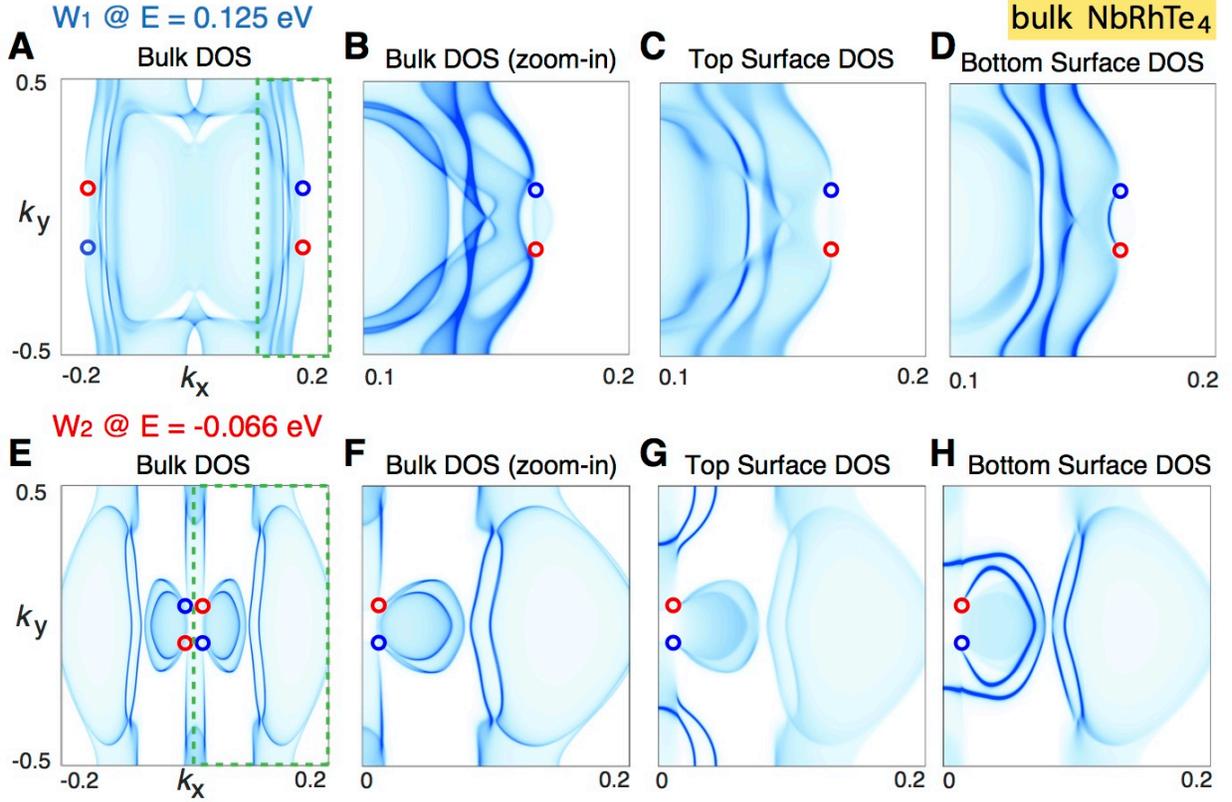

**Fig. 4. Bulk and surface density of states of bulk T$_d$-NbRhTe$_4$.** (**A,E**) Bulk DOS at the two energy levels of 0.125 eV and -0.066 eV corresponding to Weyl points W$_1$ and W$_2$, respectively. (**B,F**) Bulk DOS of the local region around the two Weyl points in the first and fourth quadrants (marked by green dashed box in a and e. (**C,G**) Surface DOS at the top surface of NbRhTe$_4$. (**D,H**) Surface DOS at the bottom surface of NbRhTe$_4$. Weyl points are marked by red and blue dots on each figure.

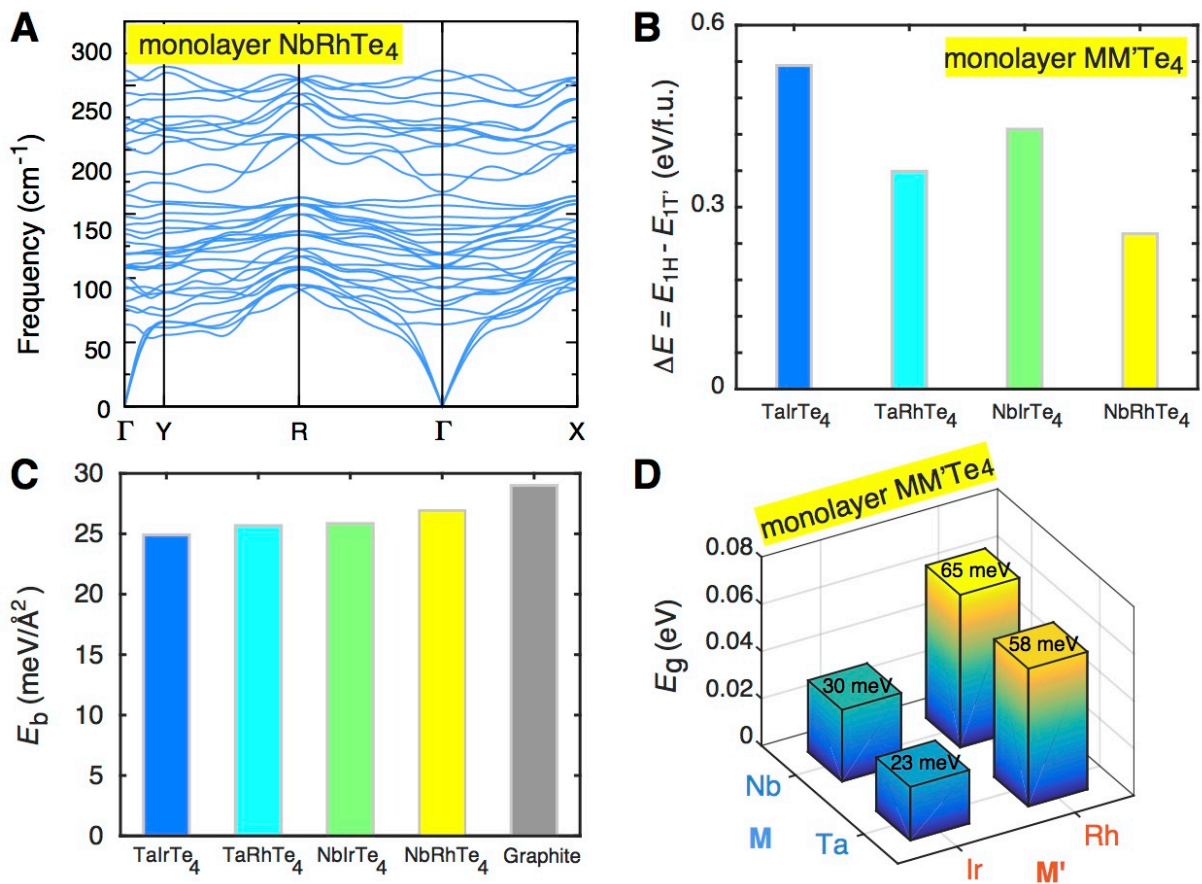

**Fig. 5. Dynamic and thermodynanic stability and electronic structure of ternary transition metal tellurides MM'Te$_4$.** (**A**) Phonon dispersion of monolayer 1T'-MM'Te$_4$ using NbRhTe$_4$ as an example. Phonon dispersions of the other three materials are presented in the Supplementary Materials. (**B**) Energy difference between the 1H and 1T' structure of monolayer MM'Te$_4$. (**C**) Interlayer binding energies of MM'Te$_4$ and graphite using the optimized optB88 exchange correlation functional for vdW interactions. Graphite is included to serve as a general reference only. (**D**) Fundamental electronic gaps of monolayer MM'Te$_4$.